# A Class of Compounds Featuring Frustrated Triangular Magnetic Lattice Cs*RE*Se$_2$ (*RE*=La - Lu): Quantum Spin-Liquid Candidates

Jie Xing[a]*, Liurukara D. Sanjeewa[a]*, Jungsoo Kim[b], G. R. Stewart[b], Mao-Hua Du[a], Fernando A. Reboredo[a], Radu Custelcean[c], Athena S. Sefat[a]

[a]Materials Science and Technology Division, Oak Ridge National Laboratory, Oak Ridge, Tennessee 37831, USA

[b]Department of Physics, University of Florida, Gainesville, Florida 32611, USA

[c]Chemical Sciences Division, Oak Ridge National Laboratory, Oak Ridge, Tennessee 37831, USA

**Abstract**

A triangular lattice selenide series of rare earths (*RE*), Cs*RE*Se$_2$, were synthesized as large single crystals using a flux growth method. This series stabilized in either trigonal (*R-3m*) or hexagonal (*P*6$_3$/*mmc*) crystal systems. Physical properties of Cs*RE*Se$_2$ were explored by magnetic susceptibility and heat capacity measurements down to 0.4 K. Antiferromagnetic interaction was observed in all magnetic compounds, while no long-range magnetic order was found, indicating the frustrated magnetism. CsDySe$_2$ presents spin freezing at 0.7 K, revealing a spin-glass state. CsCeSe$_2$ and CsYbSe$_2$ present broad peaks at 0.7 K and 1.5 K in the magnetization, respectively, suggesting the short-range interactions between magnetic rare earth ions. The lack of signature for long-range magnetic order and spin freezing down to 0.4 K in these compounds (*RE* = Ce, Yb) implies their candidacy for quantum spin liquid state.

**INTRODUCTION**

The topic of quantum spin liquid (QSL) has been attracting a great amount of interests due to its potential application for the future quantum communication and computation.[1-4] The highly entangled spins in QSL remain dynamic even at zero temperature without breaking any symmetry. The frustrated magnetic materials with degenerate ground states are excellent candidates for QSL. To date, some organic and inorganic transition metal containing triangular magnetic lattices such as κ-(BEDT-TTF)$_2$Cu$_2$(CN)$_3$, EtMe$_3$Sb[Pd(dmit)$_2$]$_2$, Ba$_3$CoSb$_2$O$_9$, Ba$_8$CoNb$_6$O$_{24}$, and NaTiO$_2$ are proposed to have interesting magnetic ground states, revealed by theoretical calculations and experimental results.[5-13] At the same time, frustrated magnetic lattices with rare-earth (RE) ions have been attractive due to the large spin-orbital coupling and anisotropic magnetic interaction.[14] Moreover, triangular magnetic lattices with RE ions exhibit diverse magnetic ground states.[15-16] For example, RE ions with an odd number of 4$f$ electrons (i.e., Kramer ions) could be treated as an effective spin $J_{eff}$ = ½. One of the famous materials is YbMgGaO$_4$ in which Yb$^{3+}$ form a triangular magnetic lattice and it was proposed by experiments and theory as a quantum spin liquid candidate.[17-21]

Recently, a large class of compounds with the formula of $AREQ_2$ ($A$ = Na, K, Rb and Cs, $Q$ = O, S, Se and Te) has been proposed as QSL candidates.[22-26] Due to the different sizes of $A$-site cations, RE ions and $Q$, $AREQ_2$ (112) is a special class of compound that tends to crystallize in different crystal space groups: LiLaO$_2$ in $P2_1/c$, LiEuO$_2$ in $Pbm$, LiYbO$_2$ in $I4_1/amd$, NaTbO$_2$ in $C2/c$, NaYbO$_2$ in $R$-3$m$, CsNdO$_2$ in $P6_3/mmc$, and NaLaS$_2$ in $Fm$-3$m$.[27-36] Among these 112 compounds, high symmetry structures maintain perfect RE triangular lattices that are separated by the A-site cations. These compounds crystallize without any disorder formation, unlike mixed occupation of Mg and Ga atoms in YbMgGaO$_4$ that may facilitate a disordered state similar to a spin liquid state.[37-39] Dzyaloshinskii-Moriya interaction is prohibited due to the magnetic rare earth ions occupying the high symmetry sites in these compounds, similar to quantum spin liquid candidate YbMgGaO$_4$.[18,40-42] Hence, 112-type compounds with triangular layers are an open ground to investigate novel frustrated magnetism.

However, among this 112 class of compounds, several important challenges remain, including limited known compounds and difficulty of growing large single crystals for finding anisotropic magnetic properties. These can be well exemplified by the Cs$RE$Se$_2$ family. So far, only the structure of CsYbSe$_2$ was reported, of which was synthesized by complex multiple steps.[43] In this work, we employed a new and simple flux method to grow large single crystals of 112 compounds, enabling anisotropic physical property measurements and inelastic neutron scattering experiments. Additionally, for the first time we confirmed that new compounds in this Cs$RE$Se$_2$ series crystallize in either trigonal or hexagonal crystal systems depending upon the size of the $RE^{3+}$ ion.

**RESULTS AND DISCUSSION**

The Cs$RE$Se$_2$ compounds were synthesized by an easy two-step method: first, the powder form of the target phase is synthesized using the stoichiometric amounts of the elements as starting materials; second, this precursor was mixed with the CsCl salt flux to obtain single crystals. More details of the synthesis procedure are given in the Supporting Information (SI). For the first time, we report that large high-quality single crystals of Cs $RE$-based 112s can be synthesized using this technique as shown in Figure 1. Additionally, we report single crystal structure characterization (details are in SI) and anisotropic magnetic properties of Cs$RE$Se$_2$ series of $RE$ = La to Lu.

We find that Cs$RE$Se$_2$ series adopt two crystal systems, trigonal $R$-3$m$ ($\alpha$-NaFeO$_2$), and hexagonal $P6_3/mmc$ ($\beta$-RbScO$_2$) types. Cs$RE$Se$_2$ with larger $RE^{3+}$ ionic radii (La, Ce, Pr, Nd, and Sm) are iso-structural and possess $\alpha$-NaFeO$_2$ structure type, while Cs$RE$Se$_2$ with smaller $RE^{3+}$ ionic radii ($RE$ = Tb, Dy, Er, Tm, Yb, and Lu) form hexagonal $\beta$-RbScO$_2$ structure type. A comparison of these two structure types is presented in Figure 2. In both structures, each of Cs, *RE* and Se atoms has a special position. In *α*-NaFeO$_2$ structure type, Cs (Wyckoff 3*b*) and *RE* (3*a*) sites have -3*m* symmetry while Se (6*c*) has 3*m* site symmetry. In comparison, in *β*-RbScO$_2$ structure type, Cs (2*c*) and *RE* (2*b*) sites are in -6*m*2 and -3*m* special positions, respectively, while Se (4*f*) is on 3*m* position. This obvious structural change among this Cs*RE*Se$_2$ family is likely due to the slight deviation of the $RE^{3+}$ ionic radii, which is supported by DFT calculation (see SI). Visually, the *α*-NaFeO$_2$ and *β*-RbScO$_2$ structure types exhibit different packing along the *c*-axis, as shown in Figure 2; herein, the atomic arrangement of one-unit cell of each of CsLaSe$_2$ and CsTbSe$_2$ is used to compare these differences. In CsLaSe$_2$ three layers of La–Se–La are packed along the *c*-axis while only two layers of Tb–Se–Tb are packed in CsTbSe$_2$. In both cases, each *RE*Se$_6$ octahedron shares edges with six surrounding units by positioning one Se atom sharing between three neighboring $RE^{3+}$ ions pointing along the *c*-axis. The polyhedral structural representation is shown in the Figure SI. 2. This interesting connectivity between *RE*Se$_6$ units creates an infinite triangular lattice of $RE^{3+}$ ions in the *ab*-plane, being displayed in Figure SI 3. Moreover, since *RE* atoms are sitting in the corners of the unit cell on *ab*-plane (Figure 2) in both structure type, therefore, the distance between $RE^{3+}$ ions is same as the size of the *a*-axis which solely depends on the size of the $RE^{3+}$ cation (Figure SI 5). Since the only structural change is that observed within the triangular layer, Cs*RE*Se$_2$ series allow understanding the role of $RE^{3+}$ ions within the same structural motifs, *RE*-ion dependent crystal electric field (CEF) anisotropy, and the exotic magnetic ground state that may be generated from the lattice frustrations.

The temperature dependence of the magnetization parallel and perpendicular to *ab*-plane above 2 K for Cs*RE*Se$_2$ (*RE* = Ce, Pr, Nd, Sm, Tb, Dy, Er, Tm, Yb) series (Figure SI 9) indicates no long-range magnetic order above 2 K. All these materials show paramagnetic behavior with antiferromagnetic interaction. Large anisotropy was found between *ab*-plane versus *c*-axis magnetization in the Cs*RE*Se$_2$. Considering the relative high values of |θ$_{CW}$|, strong frustrated phenomenon is expected in this family. A detail discussion of magnetic susceptibility, Curie-Weiss fittings, and isothermal magnetizations are given in the SI.

Moreover, we performed AC (DC 0 T, AC 2.5 Oe) and DC (0.2 T) magnetic susceptibility measurement below 2 K for the Kramer ions (Figure 3), where crystal electric field (CEF) could split the multiplet into the Kramer doublets. The slopes of DC and AC susceptibility of CsCeSe$_2$ both change at 0.7 K, while there is no difference between zero-field cooling (ZFC) and field cooling (FC), and no frequency-dependence of AC susceptibility from 1 Hz to 757 Hz. Considering the relatively high first excited energy in Ce$^{3+}$ from CEF,[44] the slope change at 0.7 K of CsCeSe$_2$ could be caused by the short-range magnetic interactions. For CsNdSe$_2$, no magnetic transition is observed in both AC and DC susceptibility down to 0.4 K. The behavior of DC susceptibility deviates from Curie Weiss law, and there is no difference between ZFC and FC. The DC susceptibility of CsSmSe$_2$ exhibits an upturn at 0.6 K. Considering the multilevel CEF for Sm$^{3+}$ and no λ anomaly at the same temperature in the heat capacity, the reason for this feature is likely low lying CEF states.[45] A magnetic transition is found in CsDySe$_2$ at 0.7 K in the AC susceptibility. The transition moves to a higher temperature by increasing frequency from 1 Hz to 757 Hz, strongly indicating the spin freezing at 0.7 K is due to the short-range interaction between Dy$^{3+}$ ions. This feature was also found in other frustrated Dy compounds.[46,47] These suggest possible spin-glass state in CsDySe$_2$. AC susceptibility of CsErSe$_2$ increases with lowering temperature down to 0.4 K. Similar behavior has been observed in the previously reported compounds, *A*ErSe$_2$ (*A* = Na and K) and ErMgGaO$_4$,[48,49] suggesting possible spin liquid ground state.

$Yb^{3+}$ triangular lattices in $CsYbSe_2$ are very important for investigating QSL due to the possible effective spin ½ of $Yb^{3+}$. The $YbSe_6$ environment in $CsYbSe_2$ is close to those in $NaYbO_2$ and $YbMgGaO_4$. It may lead $CsYbSe_2$ to be a similar $J_{eff} = 1/2$ triangular frustrated magnet.[17,23-26] Unlike $YbMgGaO_4$, $CsYbSe_2$ exhibits a broad peak at 1.5 K in magnetic susceptibility with no bifurcation in the ZFC and FC. The ground state should not be affected with CEF at such low temperature since the first excited energy from CEF in similar triangular lattice materials (such as $NaYbO_2$ or $YbMgGaO_4$) is much higher than 1.5 K.[18,24] Furthermore, no clear difference is observed in AC susceptibility measurements (1-757 Hz). Therefore, we can exclude possibility of having spin freezing in spin glass in $CsYbSe_2$. This is further supported by the heat capacity data described below.

To investigate the magnetic interaction and CEF effects, we also measured heat capacity at 0 T down to 0.4 K. Figure 4 displays a summary of the heat capacity and calculated entropy for all the compounds. The heat capacity reaches the classical limit of Dulong-Petit law for phonon heat capacity 3nR = 99.31 J/mol K (n is number of atoms per formula unit and R is gas constant).[50] Due to the full occupied electrons in the outer layers in $CsLaSe_2$ ($La^{3+}$: $4f^0$) and $CsLuSe_2$ ($Lu^{3+}$: $4f^{14}$), the heat capacity of $CsLaSe_2$ and $CsLuSe_2$ overlap and provide good phonon references for other compounds. The fitted Debye temperature for $CsLaSe_2$ is 174 K. There is no λ shape anomaly in the heat capacity indicating no long-range magnetic transition within our measuring temperature range for all compounds. Consistent with the magnetization measurement, the heat capacity results also suggest strong frustrated magnetism in these compounds. The broad peaks were observed in all the compounds except La, Pr and Lu. We also calculated the entropy from the lowest measurement temperature for each compound. The magnetic entropy of $J_{eff} = 1/2$ should be Rln2. $CsNdSe_2$, $CsDySe_2$, $CsTbSe_2$ and $CsErSe_2$ present larger values of the entropy at low temperature, which may be due to the Schottky contributions.[51] Now we focus on $CsCeSe_2$, $CsDySe_2$, and $CsYbSe_2$ that contain downturns in $M(T)$ results. Heat capacity of $CsCeSe_2$ exhibits a broad peak starting from 4 K, which is much lower than the CEF feature in the temperature dependence of the magnetization (Figure SI 9). In heat capacity, maximum of the broad peak is at 0.7 K which agrees with our magnetization data (Figure 3). However, lack of λ shaped feature indicates that it may due to short-range magnetic interaction. The magnetic entropy down to 0.4 K is 3.4 J/mol which is 60% of Rln2 in the $S = 1/2$ system. This indicates a possible residual magnetic entropy at lower temperature. Two broad peaks were exhibited in $CsDySe_2$ heat capacity. The high-temperature broad peak occurs from 3 K to 30 K whereas no significant anomaly in magnetic susceptibility in the same temperature region, indicating the origin from the Schottky contribution. The low-temperature peak reaches the maximum at 0.7 K, further confirming the spin freezing in $CsDySe_2$. In $CsYbSe_2$, there is no λ anomaly near 1.5 K as we observed in the magnetic susceptibility (Figure 3), suggesting a short-range interaction instead of the long-range magnetic order. A broad peak is observed below 10 K in $CsYbSe_2$, which is similar to $NaYbO_2$. Unlike other rare-earth ion compounds, this peak is very broad, implying the spin fluctuation in $CsYbSe_2$.

**CONCLUSION**

In this work, new $CsRESe_2$ compounds with *RE* triangular lattice were discovered, and large single crystals were for the first time grown out the salt flux. Based on the magnetization and heat capacity measurements down to 0.4 K, we found diverse magnetic states in these compounds. $CsDySe_2$ presents the spin-glass state. $CsCeSe_2$ and $CsYbSe_2$ display clear short-range interaction behavior at low temperature. Lack of long-range magnetic order and spin freezing down to 0.4 K in $CsCeSe_2$ and $CsYbSe_2$ hints their candidacy for quantum spin liquid ground state.

**Supporting Information**

The Supporting Information is available free of charge on the ACS Publications website at DOI: XXXXX

Experimental details, tables of crystallographic data, bond lengths, bond angles, temperature dependent magnetic susceptibility down to 2 K, isothermal magnetizations data and DFT calculations. (PDF)

**Accession Codes**

CCDC 1952065-1952075 contain the supplementary crystallographic data for this paper. These data can be obtained free of charge via www.ccdc.cam.ac.uk/data_request/cif, or by emailing data_request@ccdc.cam.ac.uk, or by contacting The Cambridge Crystallographic Data Centre, 12 Union Road, Cambridge CB2 1EZ, UK; fax: +44 1223 336033.

**Author Contributions**

* These authors contributed equally.

**ACKNOWLEDGEMENTS**

The research is supported by the U.S. Department of Energy (DOE), Office of Science, Basic Energy Sciences (BES), Materials Science and Engineering Division. The X-ray diffraction analysis by RC was supported by the US Department of Energy, Office of Science, Basic Energy Sciences, Chemical Sciences, Geosciences, and Biosciences Division. Work at Florida by J. S.

Kim and G. R. S. supported by the US Department of Energy, Basic Energy Sciences, contract no. DE-FG02-86ER45268.

This manuscript has been authored by UT-Battelle, LLC under Contract No. DE-AC05-00OR22725 with the U.S. Department of Energy. The United States Government retains and the publisher, by accepting the article for publication, acknowledges that the United States Government retains a non-exclusive, paid-up, irrevocable, world-wide license to publish or reproduce the published form of this manuscript, or allow others to do so, for United States Government purposes. The Department of Energy will provide public access to these results of federally sponsored research in accordance with the DOE Public Access Plan.

The authors declare no competing financial interests.

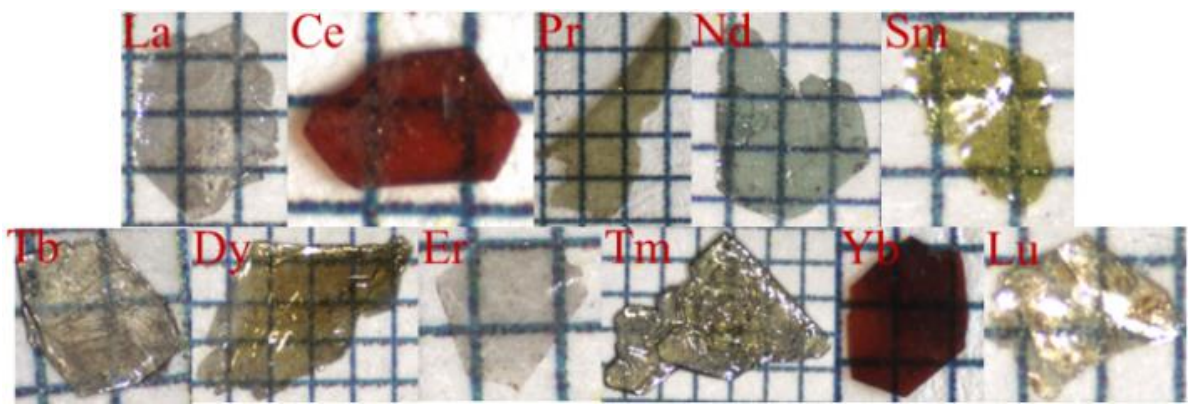

**Figure 1.** Single crystals of Cs*RE*Se$_2$ grown using salt flux. The grid is in 1 mm scale; the *c*-axis is out of the page.

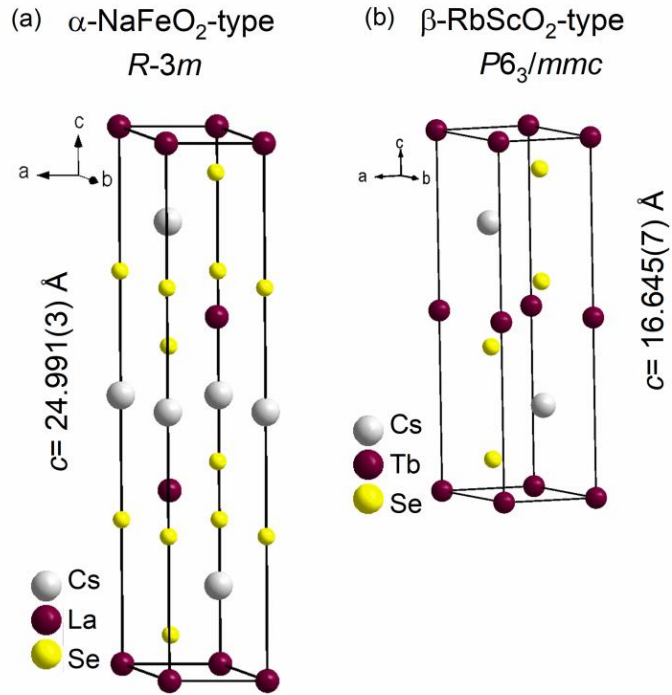

**Figure 2.** (a) Structure showing packing of Cs, La and Se atoms in CsLaSe$_2$ structure that crystallizes in *α*-NaFeO$_2$-type. (b) Packing of Cs, Tb and Se atoms in CsTbSe$_2$ structure that crystallizes in *β*-RbScO$_2$-type.

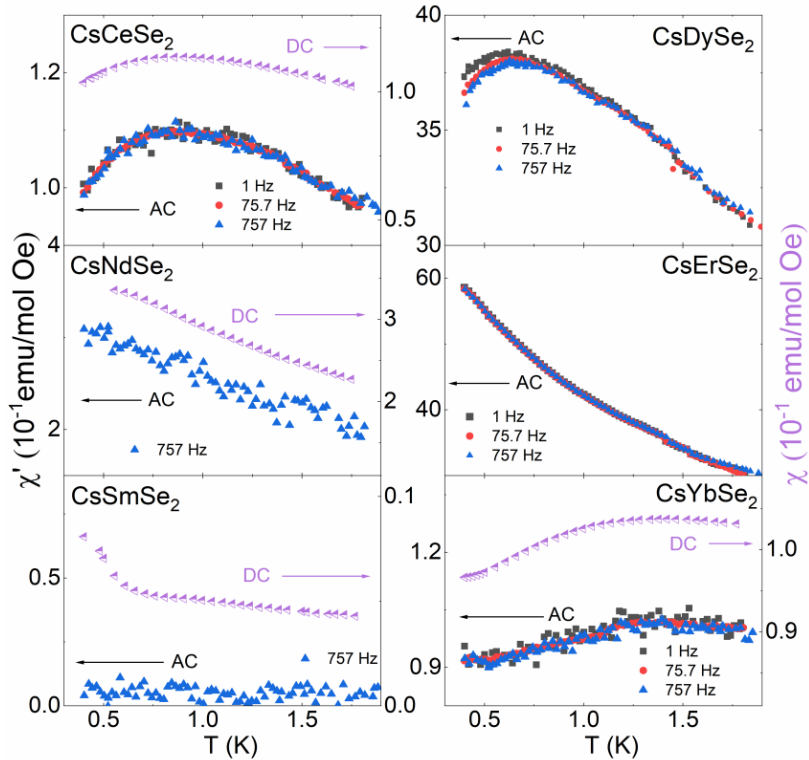

**Figure 3.** The AC and DC magnetic susceptibility below 2 K for the Kramer ions (Ce, Nd, Sm, Dy, Er and Yb).

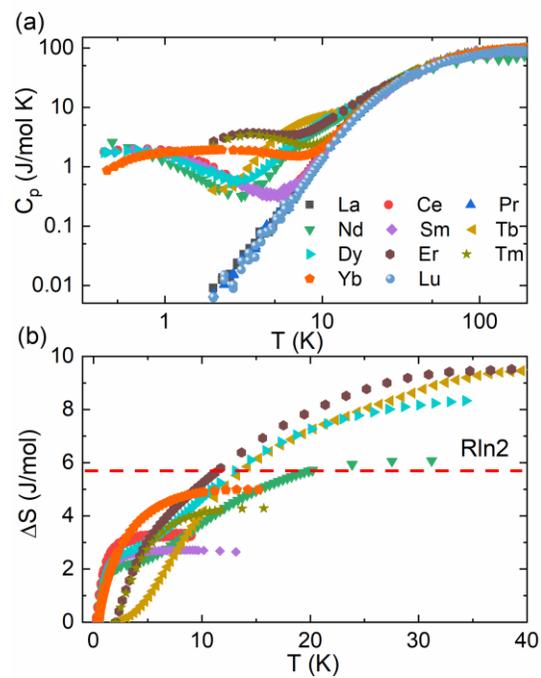

**Figure 4.** (a) Heat capacity measured at 0 T for Cs$RE$Se$_2$ series. (b) Calculated the magnetic entropy.